# RISTRETTO: Seven Spaxels Single Mode Spectrograph Design

Bruno Chazelas[a*], Christophe Lovis[a], Nicolas Blind[a], Ludovic Genolet[a], Ian Hughes[a], Michael Sordet[a], Robin Schnell[a], Anthony Carvalho[a], Maddalena Bugatti[a]

[a]Observatoire de Genève, University of Geneva, 51 chemin de Pegasi 1290 Versoix, Switzerland

**ABSTRACT**

The RISTRETTO project is aiming to build an instrument that will detect the reflected light from close-by exoplanet. It is a two stage instrument: An extreme AO system in the visible, followed by a seven spaxel single mode High resolution Spectrograph. In this paper we present the design of this spectrograph: a classical echelle spectrograph fed with single mode fibers. Standard single mode fibers have been chosen and are forming a long tilted slit in order to have the right order spacing on the detector. The instrument will be under vacuum and thermally controlled in order to make it stable.

**Keywords:** High Spectral Resolution, Single mode fiber

## 1. INTRODUCTION

RISTRETTO[1,2] is an instrument designed to detect the reflected light of an exoplanet. It is a multistage instrument starting with an extreme adaptive optics system on an 8 meter class telescope followed by a coronagraph, an IFU and a High resolution Spectrograph. The subject of this paper is to describe the design of the spectrograph. Other papers[3–5] from this conference describe the project, and the different parts composing the global instrument.

The spectrograph is designed following the heritage of HARPS and ESPRESSO. It is a high resolution echelle spectrograph. The main differences are that there are 7 input fibers and that it works at the diffraction limit. As such the fiber used for the instrument are single-mode fibers.

The spectrograph top level requirements are:
- 7 is the minimum number of fibres.
- Resolution has to be >= 130'000 goal 150'000
- Line positions should be stable on the detector at 1/100th of a pixel over 24 h
- Line Spread Function should be stable over 24h
- At a given wavelength one should be able to subtract a spectrum from an other spaxel without having LSF residual bigger than the noise (or we are able to find a way to calibrate the stellar spectrum for each spaxels).
- Sampling ~ 2.5 (> 2) pixels in the dispersion direction
- Minimum sampling in the cross dispersion direction 2.0 pixels (avoid under sampling)
- Band-pass 620 nm-840 nm (to cover the band for oxygen, water and Hα)
- The spectrograph throughput should be >= 40%

In order to achieve these requirements, one needs in addition to have the instrument in vacuum and thermally controlled. The detector is a deeply depleted standard E2V CCD231-84-x-E74, with an astro-multi-2 coating.

## 2. OPTICAL DESIGN

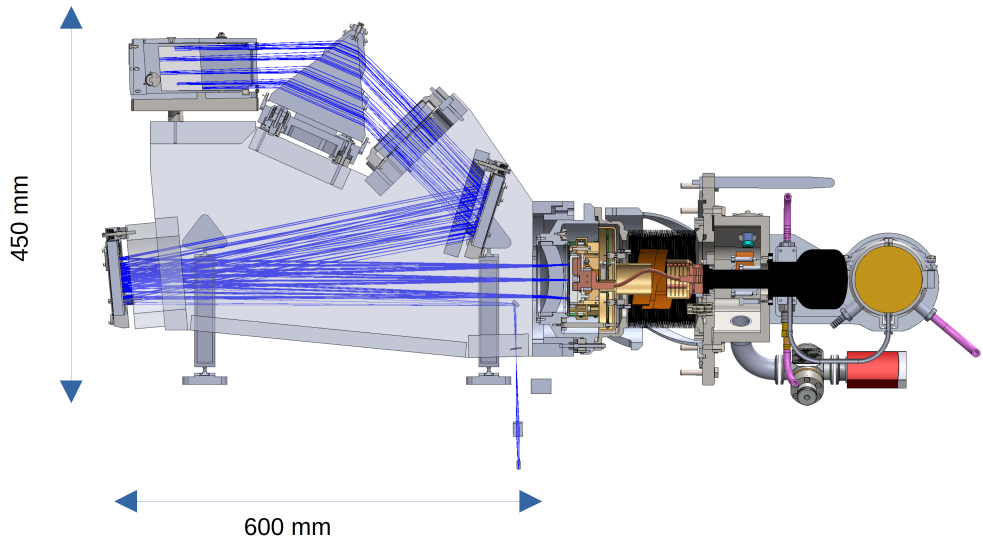

Figure 1: Optical Design of the spectrograph. A double pass Echelle spectrograph, cross dispersed by a prism.

Given the specification an existing grating from MKS Richardson grating company has been chosen as the main dispersive element for the spectrograph. It is an R2 grating (63 degree blaze angle), with 23.2 lines per millimetre. It gives a target resolution of 135000 with a sampling of 2.5 pixel (see figure 2). The grating size, with respect to the standard diffraction limit formula has been increased by a factor 1.7, in order to preserve the resolution[6]. With this choice we essentially have an instrument where beams are propagated as Gaussian beams (with a low level of aberrations). Most of the other optics are lenses. Mirrors have been used to make the instrument more compact.

The design inspired by B. Delabre, is a double pass "white pupil" spectrograph, with a prism cross disperser.

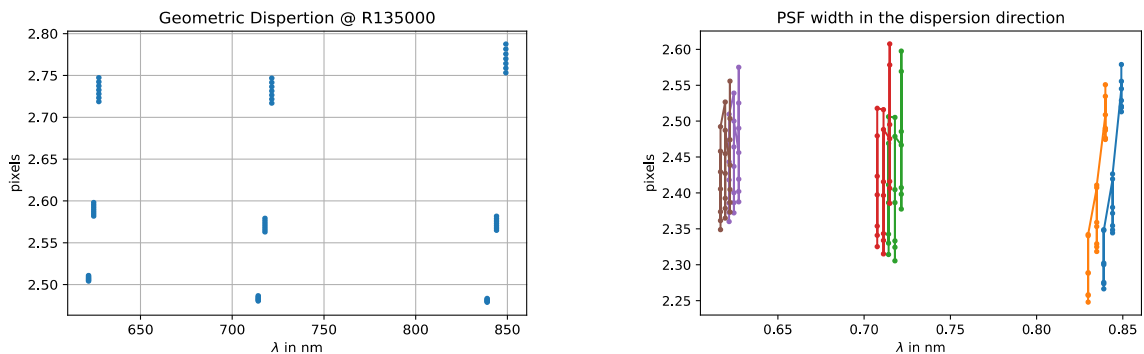

Figure 2: Dispersion of the spectrograph and size of the PSF across the detector and for different fibers.

One particular difficulty: in a cross dispersed echelle spectrograph the spacing of the different fiber orders on the detector derives from the distance between the input fibers and in our case the optimal distance is around 25 microns which is smaller than the typical fiber size. The strategy we have adopted is to still use standard single-mode fiber, but to align them on a long slit that is nearly horizontal. Tuning the angle of this slit allows to set the fiber order distance on the detector precisely at desired distance. The disadvantage of this configuration is that the different fiber orders are offset on the detector in the dispersion direction. Thus efficiency of the different fiber orders will be varying differently across the spectrum. On the red and blue end wavelength the cut will also be different for the different fibers. This has however an advantage for two different fibers the same wavelengths we'll be separated 5 times more, which will help to avoid coherent interferences.

This choice has also consequences on the IFU design this has been described in a paper[4] in the same conference.

A thermal analysis (see figure 3) of the optical design has been made and showed that the thermal sensibility needed for the instrument is about 20 mK.

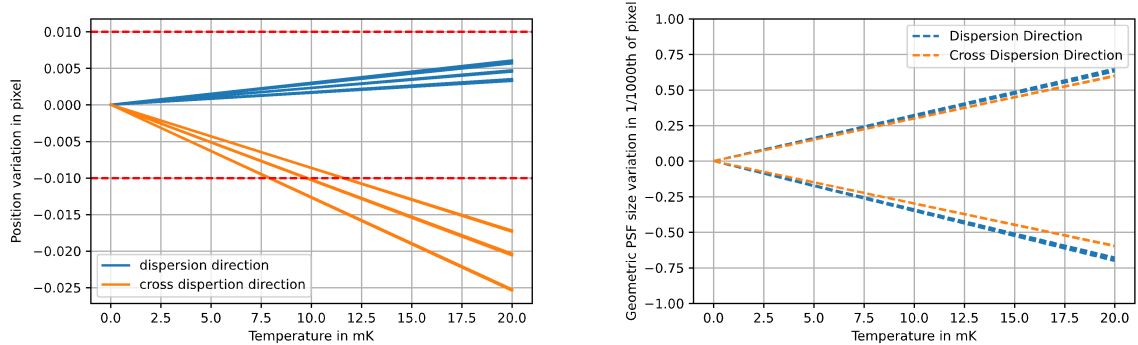

Figure 3: Thermal sensitivity of the optical design. Below 20 mK of temperature variation the shift of the PSF is below 0.01 pixel in the dispersion direction, and ~ 2 times more in the cross dispersion direction. The focus variation are also negligible.

Being diffraction limited the PSF size is mainly driven by the wavelengths in order to have more even sampling (see figure 2) across the whole spectrum the solution that has been chosen is focus and tilt the detector so as to get a more uniform PSF size.

## 3. EXPOSURE METER

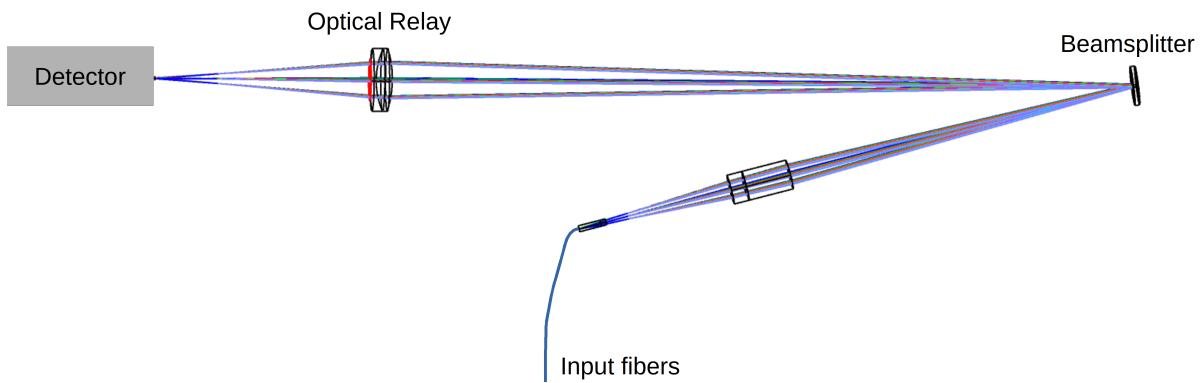

Figure 4: Scheme of the spectrograph exposure meter, note that the fibers lenses and beamsplitter are located inside the vacuum chamber while the detector is outside.

It is necessary for the instrument to measure in real time the flux of the different input fiber. in order to do that a beamsplitter is positioned near to an intermediate focus and the white image of the fiber are made onto a detector. The coefficient of reflection of the faces of the beamsplitter are bellow 1% thus the ghost on the main spectrum is negligible. The detector will be situated outside the vacuum tank and is a commercially available CMOS detector. Given the performance of the detector and of the system we can expect to have flex measurement our star like Proxima B with a speed of about 1 Hertz.

## 4. CALIBRATION SYSTEM

In order to calibrate the spectrograph a calibration system has been designed very similar to the calibration systems used for harps or espresso. One needs a white light source for flat fielding the spectrograph. A super continuum laser will provide the white light. The wavelength reference will be a uranium neon lamp. A single mode Fabry-Perot is being designed. The whole system is using single-mode fiber and the switching between the different sources is done using commercially available single-mode fiber switchers.

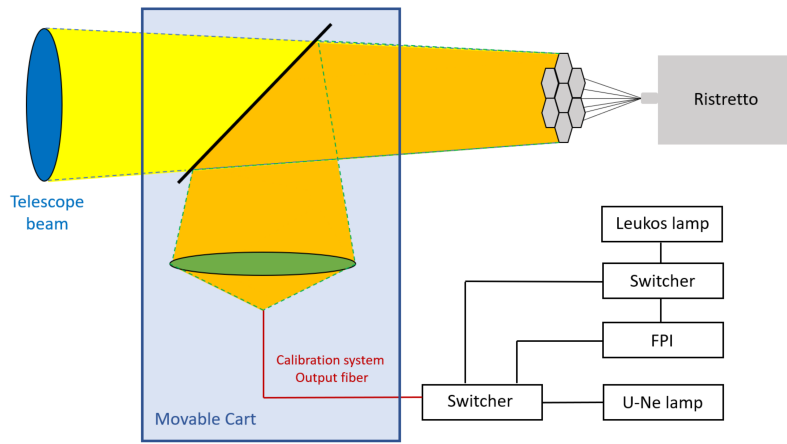

Figure 5: Scheme of the RISTRETTO spectrograph calibration system (From M. Bugatti Master Thesis)

## 5. OPTO-MECHANICAL DESIGN

Given the nature of the instrument, the stability requirements ~9 m.s$^{-1}$ are relaxed compared to HARPS or ESPRESSO (1 m.s$^{-1}$/ 10 cm.s$^{-1}$). Thus the choice has been made to use aluminum has the material for the optical bench and the optical mounts. We however follow these principles to keep the instrument stable: there are no moving part inside the instrument; all the necessary adjustments are done during the initial alignment mainly using shims. The grating and the prism are linked to their mounts using metal pads that are glued directly on the optics in a similarly to what has been done for HARPS or ESPRESSO.

## 6. THERMAL INSULATION DESIGN

There are two major constraints in this instrument:

1. The optics of the instrument have to remain stable within 20 mK. This has to be true on different timescales during the night and on several months.
2. The spectrograph will be operated on the nasmyth platform of an 8 meter class telescope where an extreme adaptive optics system will be operated. it is thus extremely important that the spectrograph does not radiate significant power and follows the ambient temperature so as not to produce additional turbulence. As we plan to propose the instrument as a visitor instrument for the VLT, we follow the ESO requirements[7] for the thermal perturbation: "the maximum temperature difference between any exposed surface of the instrument (or of any associated equipment) and ambient shall be ≤ +1.5/-5.0 ∘ C in wind-still conditions, with a maximum upwardly convected energy for the instrument and all associated equipment of 150 W".

In order to achieve all these goals we have made the following design (see figure 6):

The spectrograph is in a vacuum chamber, In the chamber there are a radiation shield and heaters that will be regulated. The feet of the tank are insulated from the holding structure and also have heaters included to "cut" the seasonal temperature variations. The chamber itself is in a box that is maintained at constant temperature of ~18°C. The box is insulated and the chosen temperature allow to meet the tight thermal requirement for an instrument on a Nasmyth platform. The box is thermally controlled by a system based on a cold glycol source, a glycol/air heat exchanger, heaters and a regulation system.

Control Racks thermal regulation is provided by ESO standard regulation system (also based on a cold glycol source).

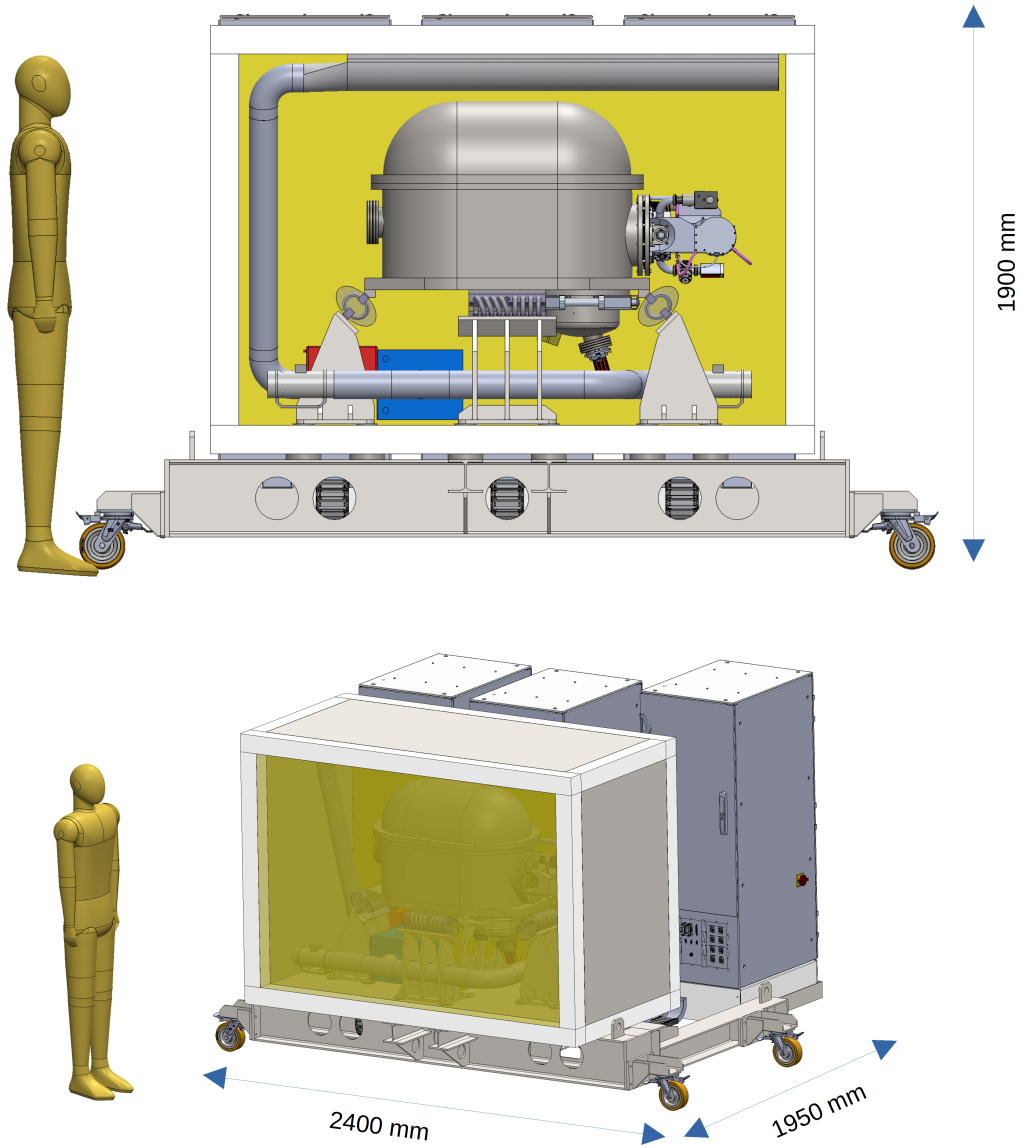

Figure 6: RISTRETTO spectrograph inside its thermal enclosure and with its needed control electronic racks.

# 7. DETECTOR HEAD DESIGN

Similarly to HARPS and ESPRESSO the instrument is operated at ambient temperature while detector required to work at 158 K. The solution is to run the detector inside a differential vacuum cryostat. For RISTRETTO we have designed a system inspired by previous designs with a major change being that the cooling is provided by a cryocooler.

The chosen cryocooler is from Thales, model LPT9310. It is provided with a vibration reducing system. It has been chosen for its reliability (it is space qualified) and because we used it successfully on a cryostat for the same CCD that is mounted on the EULER telescope. As one of our main concerns was vibrations, We performed a combined FEM and optical analysis of the system and concluded that vibration would enlarge the PSF in a negligible manner and would have no impact on the positions of the lines.

Shown on the figure 7, the cool head containing the detector is rigidly fixed to the optical bench a flexible Bellow connect this cold part to the vacuum tank shell maintaining a differential vacuum

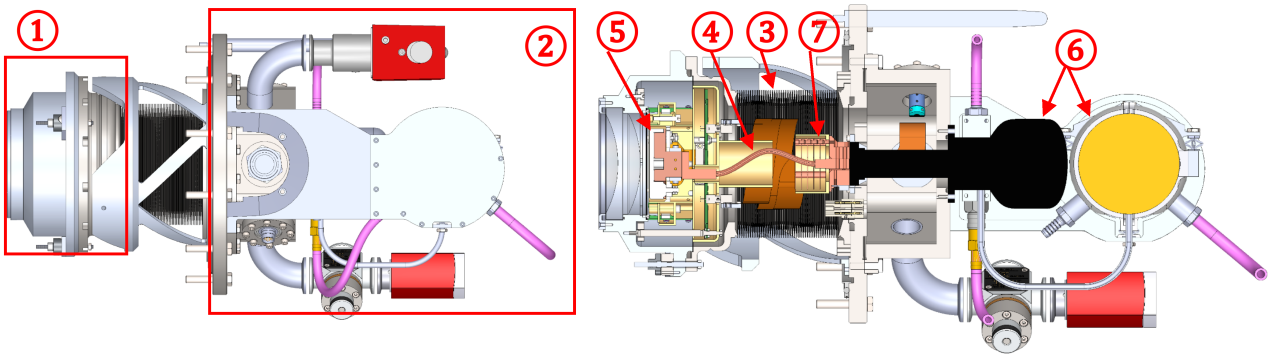

Figure 7: RISTRETTO spectrograph Detector head design. he detector unit consists of a detector head subassembly ① and a cryocooler subassembly ②. the connection between the 2 is ensured by flexible components (bellow ③ and thermal strap ④). The detector head and the cryocooler subassemblies are hard mounted respectively on the optical bench and the vacuum chamber. The detector unit has its own evacuated enclosure. The detector unit regulates the temperature of the detector ⑤ at 158K using a cryocooler ⑥ and heaters. At the cold tip of the cryocooler, a cryopump ⑦ made of activated charcoal is implemented to maintain a low pressure without active pumping.

# 8. CONCLUSION

In this paper we have presented the design of the RISTRETTO spectrograph. The design phased is nearly finished and manufacturing will start soon. When it will be ready it will undergo tests on sky on the EULER telescope paired with the KalAO[8] adaptive optic system.

# 9. ACKNOWLEDGMENTS

We would like to thanks in particular Bernard Delabre that helped us with the spectrograph design concept.

This work has been carried out within the framework of the National Centre of Competence in Research PlanetS supported by the Swiss National Science Foundation under grants 51NF40_182901 and 51NF40_205606. The RISTRETTO project was partially funded through SNSF FLARE program for large infrastructures under grants 20FL21_173604 and 20FL20_186177. The authors acknowledge the financial support of the SNSF.

# 10. REFERENCES

[1] Lovis, C., Snellen, I., Mouillet, D., Pepe, F., Wildi, F., Astudillo-Defru, N., Beuzit, J.-L., Bonfils, X., Cheetham, A., Conod, U., Delfosse, X., Ehrenreich, D., Figueira, P., Forveille, T., Martins, J. H. C., Quanz, S. P., Santos, N. C., Schmid, H.-M., Ségransan, D., et al., "Atmospheric characterization of Proxima b by coupling the SPHERE high-contrast imager to the ESPRESSO spectrograph," Astronomy & Astrophysics **599**, A16 (2017).

[2] Chazelas et al., "RISTRETTO: a pathfinder instrument for exoplanet atmosphere characterization," December 2020, SPIE.

[3] Lovis, C. et al., "RISTRETTO: high-resolution spectroscopy at the diffraction limit of the VLT | SPIE Astronomical Telescopes + Instrumentation," Ground-based and Airborne Instrumentation for Astronomy IX **12184**, SPIE (2022).

[4] Kühn, J. G. et al., "RISTRETTO: A high-throughput high-contrast coronagraphic IFU at two diffraction beam widths," Advances in Optical and Mechanical Technologies for Telescopes and Instrumentation V **12188**, SPIE (2022).

[5] Blind, N. et al., "RISTRETTO: coronagraph and AO designs enabling High Dispersion Coronagraphy at 2 lambda/D," Adaptive Optics Systems VIII **12185**, SPIE (2022).

[6] Robertson, J. G. and Bland-Hawthorn, J., "Compact high-resolution spectrographs for large and extremely large telescopes: using the diffraction limit," 5 October 2012, Amsterdam, Netherlands, 844623.

[7] "Requirements for Scientific Instruments on the VLT Unit Telescopes.", VLT-SPE-ESO-10000-2723, ESO (2005).

[8] Hagelberg, J., Restori, N., Wildi, F., Chazelas, B., Baranec, C., Guyon, O., Genolet, L., Sordet, M. and Riddle, R., "KalAO the swift adaptive optics imager on the 1.2m Euler Swiss telescope in La Silla, Chile," Adaptive Optics Systems VII **11448**, 1460–1467, SPIE (2020).